\documentstyle[epsf]{mn}

\newif\ifAMStwofonts

\def\etal{\sl et al.~\rm}
\def\deg{\mbox{${}^\circ$}}

\def\apj{ApJ}

\def\db{{\bf d}}

\def\Ab{{\bf A}}
\def\Bb{{\bf B}}
\def\Cb{{\bf C}}
\def\Db{{\mathbf \Delta}} 
\def\Eb{{\bf E}}
\def\Fb{{\bf F}} 
\def\Ib{{\bf I}}
\def\Mb{{\bf M}}

\def\Tb{{\bf T}}
\def\Wb{{\bf W}}

\def\Sb{{\mathbf \Sigma}}
\def\Ksc{{\it K}}
\def\Lsc{{\it L}}
\def\Msc{{\it M}}

\title[Applications of Wavelets to CMB Maps]
  {Applications of Wavelets to the Analysis of Cosmic Microwave
Background Maps}
\author[L. Tenorio et al.]
  {L.~Tenorio,$^1$ 
  A.H.~Jaffe,$^2$ S.~Hanany,$^{2,3}$ C.H.~Lineweaver$^4$ \\
  $^1$Mathematical and Computer Sciences, Colorado School of Mines,
      Golden, CO 80401\\
  $^2$Center for Particle Astrophysics,
      University of California, Berkeley, CA 94720\\
  $^3$School of Physics and Astronomy, University of 
        Minnesota, Minneapolis, MN 55455\\
  $^4$School of Physics, University of New South Wales, Sydney Australia}

\def\LaTeX{L\kern-.36em\raise.3ex\hbox{a}\kern-.15em
    T\kern-.1667em\lower.7ex\hbox{E}\kern-.125emX}

\begin{document}

\label{firstpage}

\maketitle 

\begin{abstract} 

We consider wavelets as a tool to perform a variety of tasks in the
context of analyzing cosmic microwave background (CMB) maps.  Using
Spherical Haar Wavelets we define a position and
angular-scale-dependent measure of power that can be used to assess
the existence of spatial structure.  We apply planar Daubechies
wavelets for the identification and removal of points sources from
small sections of sky maps.  Our technique can successfully identify
virtually all point sources which are above $3\sigma$ and more
than 80\% of those above $1\sigma$. We discuss the trade-offs between
the levels of correct and false detections.  We denoise and compress a
100,000 pixel CMB map by a factor of $\sim 10$ in 5 seconds achieving
a noise reduction of about 35\%. In contrast to Wiener filtering the
compression process is model independent and very fast.  We discuss
the usefulness of wavelets for power spectrum and cosmological
parameter estimation.  We conclude that at present wavelet
functions are most suitable for identifying localized sources.

\end{abstract}


\section{Introduction}

The cosmic microwave background (CMB) radiation encodes a vast amount of
information about the early universe and its subsequent evolution.  A
number of experiments: ground-, balloon- and space-based, are poised to
generate large data sets from which one hopes to decode this
information. Optimal data analysis methods to be utilized with these
data sets are presently an active area of research.  In this paper we
explore the application of wavelet methods 
to the analysis of CMB data sets. 

Cosmological theories usually model the CMB sky as a homogeneous
random field on the sphere. Spherical harmonics arise naturally
through the spectral decomposition of the field. They are optimal for
frequency localization---indeed, they define spatial frequencies,
$\ell$, on the sphere---but they do not allow assessment of local
features. To represent a signal that is nonzero
only on a small patch of the sky a large number (formally infinite) of
spherical harmonics are required to obtain all necessary cancellations
outside the patch.  Wavelets, on the other hand, are locally supported
and therefore, only those supported in the patch are needed to
represent the signal.

Wavelets have become an attractive tool for data analysis and
compression because they are computationaly efficient and have better
time-frequency localization than the usual Fourier methods. Although
wavelets are usually defined on the real line, or subsets of $R^n$,
there have been some recent generalizations to other surfaces like
the sphere. In this paper we use spherical wavelets when dealing with
large portions of the sky and planar wavelets for small patches.

Spherical Haar wavelets (SHW) were introduced by Sweldens (1995) as a
generalization of planar Haar wavelets to the pixelized unit sphere.
The basic idea of SHW is to transform the original map into the sum of a
smoothed lower resolution map and a set of coefficients
encoding the small-angular-scale details not
captured by the smoothed map. The smoothing is done by lowering the
resolution of the map through a hierarchical pixelization scheme.
Thus, pixelizations of sky maps automatically yield SHW, and these are
useful to study data as a function of position and angular scale.  SHW
are not smooth but are nonetheless attractive because they can be
easily implemented and used as a computationally efficient
exploratory data analysis tool.  In Section \ref{secrms} we
present a wavelet formalism on the sphere adapted to CMB sky maps. We 
then use this formalism to compare structure in the  COBE-DMR maps
with predictions of expected structure based on model dependent 
simulations.

A common thread throughout the rest of the paper  
is that of denoising
a noisy signal. Because wavelet functions have excellent localization
properties and their `thresholding' is the asymptotically best way to
remove additive Gaussian white noise (Donoho 1992), they are
useful for both the identification of sources embedded in Gaussian
noise and for suppression of such noise. A further attractive feature
of wavelet denoising is that the process is computationally
efficient. In Section \ref{subloc} we use wavelets for the
identification of point sources embedded in noise.  We do not use SHW
for this particular application because SHW are not 
smooth and thus are not optimal for
estimating smooth localized sources. We do not know of any
computationally efficient smooth wavelets on the sphere. For these
reasons the algorithm we propose is more appropriate for small
patches of the sky (where the flat sky approximation is appropriate)
and we use tensor products of one dimensional Daubechies wavelets.
Wavelet decomposition of CMB maps for the purpose of
identifying non-Gaussianity has been considered recently by Ferreira,
Magueijo \& Silk (1997), Pando, Valls-Gabaud \& Fang (1998), and
Hobson, Jones \& Lasenby (1998). Several other groups are
concentrating efforts on usage of wavelet for foreground
identification and removal (e.g., Cayon 1999, Sanz \etal 1999).

Current methods for compressing and denoising CMB maps are based on
Karhunen-Loeve transformations (Bond 1995, Bunn \& Sugiyama 1995) or Wiener
filtering \cite{bunn+96}.  For a sky map with $N$ pixels the
computational cost of these methods is $O(N^3)$ so their brute force
application to future mega-pixel maps ($N\approx 10^6$) is an unsolved
(and perhaps unsolvable) computational challenge.  Map denoising by
wavelet thresholding is significantly more efficient taking only
$O(N{\rm log}(N))$ operations.  For example, applying the wavelet
transform, denoising and reconstructing a $\approx 10^5$ pixel map
takes less than 6 seconds on an Alpha 500/266. Furthermore, wavelet
thresholding can be done in a model independent way.  In Section
\ref{secmap} we present such a technique using SHW.

It has been pointed out repeatedly  
(e.g., Bond, Jaffe \& Knox 1998, Borrill 1999,
Bond \etal 1999 and references therein) that
because of the size of future data sets new analysis techniques will
need to be developed.  Techniques that can efficiently extract
cosmological information from mega-pixel sky maps. In Section
\ref{secparam} we discuss the usefulness of wavelets for this
task.  Section \ref{secsum} summarizes the results. A summary of SHW
applied to the COBE pixelization can be found in the Appendix.

\section{Wavelet Power and CMB Sky Maps}
\label{secrms}

A CMB map is a vector $\Tb=(T_i)$ of temperatures $T$ in pixels
$i$, located at position ${\hat x_i}$.  We decompose this
vector---which is also a field on the sky---into its
spherical-harmonic components,
\begin{equation}
    T({\hat x})=\sum_{\ell m} a_{\ell m}Y_{\ell m}({\hat x}).
    \label{eq:alm}
\end{equation}
We can then define the power spectrum in the usual way as the
expectation of the square of the components
\begin{equation}
    \langle a_{\ell m}a_{\ell' 
    m'}\rangle=C_{\ell}\delta_{\ell\ell'}\delta_{mm'}.
    \label{eq:Cldef}
\end{equation}
Up to normalizing constants, the expected value of the RMS$^2$ of the
map is
\begin{equation}
\langle\sum_i
T_i^2\rangle   \propto  \sum_{\ell}(2\ell+1)C_{\ell}/{4\pi}. 
\end{equation}
An observation, $d_i$, of this (pure-signal) map is the sum of the map
with some noise component, $d_i=T_i+n_i$, so
\begin{equation}
  \langle\sum_i d_i^2\rangle   \propto  \sum_{\ell}(2\ell+1)C_{\ell}/{4\pi}+N_{ii'}
\end{equation}
where $N_{ii'}$ is the noise correlation matrix, and in particular
$N_{ii}=\sigma_i^2$ is the noise variance of pixel $i$.

We now introduce an analogous formalism, expressing the map RMS$^2$
using wavelets. Since wavelets are localized both in the spatial and
frequency domain, a wavelet expansion easily leads to partitioning of
the RMS$^2$ into components from different angular scales and from
distinct portions of the map. We introduce the formalism in terms of
spherical wavelets, most suitable for maps covering large portions of
the sky.

Spherical Haar wavelets were introduced by Schroeder \& Sweldens
(1995) as an example of what they call second-generation wavelets. These
are wavelets which are not dilations and translations of a mother wavelet
and which can be defined in spaces more general than $R^n$.  As far as
we know they were the first to introduce computationally
efficient wavelets on the sphere. Sweldens (1995) applied
SHW and other
second-generation spherical wavelets to data compression.

The SHW transform of a map
\[\db=(d_i)=\vec{\lambda}_J\]
 of maximum resolution $J$ is 
(see Appendix)
\[
\Wb\db= \vec{\gamma} = (\vec{\lambda}_{J_o}, \vec{\gamma}_{J_o},\cdots,
\vec{\gamma}_{J-1})^t,
\]
where the vector $\vec{\lambda}_{J_o}$ is the map at the lowest
resolution, $J_o$, and $\vec{\gamma}_{j}$, $J_o \leq j \leq J$, is a
vector of wavelet coefficients at scale $j$.  We define the scale-$j$
power in the wavelet domain as a normalized sum of squares of the
coefficients at scale $j$.  To interpret this measure we show its
relation to the RMS$^2$ of a map:
\(
{\rm RMS}^2 =  \sum_i w_i^2   d_i^2 / \sum_i w_i^2,    
\)
for chosen weights $w_i$.
The wavelet transform
(where the weights $w_i$ are implicitely included in $\vec{\lambda}_J$) yields 
an angular-scale decomposition of the RMS$^2$
\begin{eqnarray}
 {\rm RMS}^2  & = &
\frac{1}{w} (\vec{\lambda}_J )^t \vec{\lambda}_J \nonumber \\
 & = & \frac{1}{w}[\,
(\vec{\lambda}_{J_o} )^t  \Ab_{J_o}\vec{\lambda}_{J_o}  + 
(\vec{\gamma}_{J_o} )^t   \Bb_{J_o}\vec{\gamma}_{J_o} \nonumber \\
 &  & + \cdots +
(\vec{\gamma}_{J-1} )^t   \Bb_{J-1}\vec{\gamma}_{J-1} \,] \nonumber \\
 & = & {\rm RMS}^2_{J_o} + \cdots + {\rm RMS}^2_{J}, 
\label{rmsdec}
\end{eqnarray}
where $w=\sum_i\sqrt{ w_i^2 }$ and $\Ab_{J_o},
\Bb_{J_o},...,\Bb_{J-1}$ are diagonal matrices defined by the wavelet
transform (see Appendix).  Expression (\ref{rmsdec}) separates
contributions to the RMS$^2$ from different resolutions. Each ${\rm
RMS}^2_{i}$ corresponds to a window function of $C_\ell$. Table 1
shows the results of the RMS$^2$ decomposition of some spherical
harmonics of different order $\ell$. The table shows the percentage of
the total RMS$^2$ that goes into each scale. As expected, as $\ell$
increases a higher percentage of the power goes into the higher scales
(i.e., higher $j$).  Table 1 also shows that spherical harmonics are
not well localized in wavelet scale.

\begin{table}
\caption{Percentage of the total RMS$^2$ that goes into each scale $j$
for some spherical
harmonics of different orders $\ell$. The lowest and highest
resolutions are $J_o=5$ and $J=8$, respectively.}
\label{tbl1}
\begin{tabular}{@{}ccccc} 
 $\ell$  &  RMS$^2_5$   & RMS$^2_6$   &  RMS$^2_7$   & RMS$^2_8$ (\%) \\  \hline
  2      &   100                 &   0          &   0           &    0  \\
   20    &      73      &       20      &       6       &       1 \\
   60   &       3       &       53      &       32      &       12 \\
  500   &       2       &        6      &       21      &       71 \\   
\end{tabular} 
\end{table} 

It is useful to construct the standardized sum of squares, 
SSQ$_j =(\Db \vec{\gamma}_j )^t \Db\vec{\gamma}_j $,
where $\Db$ is a diagonal matrix that transforms the covariance matrix
of the SHW coefficients to the identity.
Assuming independent Gaussian
noise the null distribution of SSQ$_j$, i.e., its distribution 
in the absence of signal, is a $\chi^2_{k_j}\approx
N(k_j,2k_j)$, where $k_j$ is the subspace dimension at scale $j$. 
To search for structure in the map at scale $j$ we use the statistic 
\begin{equation}
z_j=\frac{{\rm SSQ}_j-k_j}{\sqrt{2 k_j}}.
\label{zeq}
\end{equation}
With uncorrelated noise and known noise RMS the $z_j$ are centered 
at zero if and only if  
there is no structure at scale $j$, regardless of the   
Gaussianity of the noise.
Also, if there is no structure then, by the central limit
theorem, the $z_j$ are approximately Gaussian-distributed even
with uncorrelated non-Gaussian noise.     
The distribution of $z_j$ is thus an indication of structure.
 
\begin{figure} 
\caption{Histograms of the $z_j$ statistics  
for 1000 simulations of beam smoothed, constant power 
($\ell(\ell+1)C_{\ell}=$ constant)
large-angular-scale maps, $\ell \leq 28$. 
The dashed lines correspond to the values of  
$z_j$ for the actual DMR 53+90 GHz DMR.}
\centerline{\epsfxsize=8cm\epsfbox{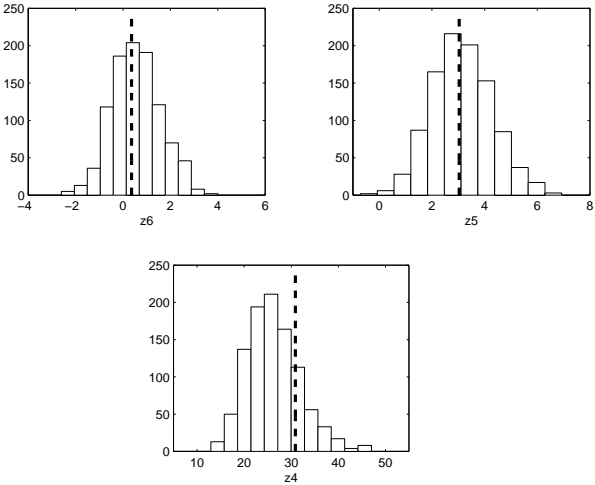}} 
\label{dmrzs}
\end{figure}

We apply our statistic to test for structure in the COBE DMR
map. Figure \ref{dmrzs} shows histograms of the $z_j$
(Eq. \ref{zeq}) for 1000 simulations of large-angular-scale maps  
$\ell \leq 28$, with $\ell(\ell+1)C_{\ell}=(24\pi/5)Q^2_{\rm rms-PS}$, 
with $Q_{\rm rms-PS}= 15.3\,\mu$K, $7\deg$ FWHM beam smoothing
and the noise level of the DMR 53+90 GHz map. The dashed lines in the
figure correspond to the actual values of $z_j$ for the DMR 53+90 GHz
map. The values are consistent with those expected from the
simulations. Both the simulations and the DMR maps have a $20\deg$
galactic cut.  SHW are easily adaptable to sky maps with Galactic cuts
or with any others kinds of gaps.  The wavelet functions are still
orthogonal and the null distributions are exactly as before (with
smaller $k_j$). We note that with only a $10\deg$ galactic cut 
(i.e., with considerable galactic contamination) the
values of the $z_j$ obtained from the DMR maps are 71.2, 5.9, and 1.6
for $j=4$, 5 and 6, respectively. The histograms representing the
simulations, which have only a CMB signal in them, remain
virtually unchanged.  As expected, more significant structure is    
detected when some of the galactic signal is present, and this
structure is not consistent with a cosmological signal alone.  

At resolution 4 the mean of the simulations and the actual map are
both considerably displaced from zero. At resolution 5 $z_5$ is
somewhat less displaced from zero and at resolution 6 $z_6$ is
consistent with zero. This is consistent with the beam scale being
right around the pixel size at resolution 5 so there is    
structure on that scale and 
at the lower resolution, but no structure at the higher resolution.

As an aside, note that the covariance matrix
$\Sb_\gamma=\langle\vec{\gamma}\vec{\gamma}\rangle$ of
$\vec{\gamma}$ is
\begin{equation}
\Sb_\gamma=\Wb(\Cb  + \Sb_{\rm noise})\Wb^t, 
\label{covgamma}
\end{equation}
where $\Cb$ and $\Sb_{\rm noise}$ are the covariance matrices of
the model and noise, respectively. 
In general the matrix $\Db$, which is used to transform the covariance 
matrix of $\vec{\lambda}$ to the identity, is
\begin{equation}
\Db=\Sb_\gamma^{-1/2}.
\label{Deqn}
\end{equation}
If we take no correlations from
a cosmological model ($\Cb=0$) and the noise is uncorrelated and
homoscedastic  
(i.e., $\Sb_{\rm noise}\propto \Ib$), then $\Sb_\gamma$ is diagonal. 
But $\Sb_\gamma$ is a lot more complicated when a
cosmological model is assumed.   

Our analysis has not taken full advantage of the localization
properties of wavelets. SHW can easily yield a spatial decomposition.
First, divide the sky into $P$ disjoint patches. Each patch $i$ can be
represented as $\vec{\lambda}_{J,i}$ by filling with zeros those pixels
not in patch $i$. All the $\vec{\lambda}_{J,i}$ are orthogonal and
\[{\rm RMS}^2 = \frac{1}{w}[(\vec{\lambda}_{J,1} )^t \vec{\lambda}_{J,1} 
+ \cdots + (\vec{\lambda}_{J,P} )^t \vec{\lambda}_{J,P}] .  \]
An angular-scale decomposition like (\ref{rmsdec})
can now be applied to every term.

\section{Wavelet Denoising of a Map}
\label{secden}

In the following sections we apply wavelet denoising with two
different applications in mind. In the first application denoising will be
used to identify and remove localized sources from a CMB map. In
the second application denoising will be used to suppress instrumental,
or other non-localized noise, and compress a map. In both
operations we use the same denoising technique. 
The techniques discussed are applicable to both spherical and 
flat space wavelets. Both types of wavelets will be considered.

Let $d_i$ be the measured temperature in pixel 
$i$, we write
$d_i = T_{s,i} + F_i + n_i$, where
$T_{s,i}$ is the underlying 
temperature at pixel $i$, $F_i$ is a foreground source temperature, 
and $n_i$ is a noise term describing noise 
sources which are not localized on
the sky (e.g., instrumental, atmospheric ).
The assumptions we make
about $T_{s,i}$ depend on whether we remove sources or suppress noise in
a map. In the first case 
we assume that $T_{s,i}$ and $n_i$ have zero mean
and are uncorrelated, and that $F_i$ is fixed and unknown. $T_{s,i}$ is
assumed random, drawn from some power spectrum, 
and we attempt to  discriminate fixed 
sources from random realizations of ${\Tb}_s$ in our sky. 
In the second application, when suppressing noise in 
a map, both $T_{s,i}$ and $F_i$ are assumed fixed. 
The goal is to suppress $n_i$ and thus obtain a better estimate of the
actual realization of the signal in our sky.  

In either case the procedure consists of `denoising' by thresholding
the wavelet coefficients of the data and transforming back.  The
difference between the two cases arises in what is assumed as noise 
during the thresholding stage. This will be discussed in
Section \ref{section noise}.  We start by computing
the wavelet transform of the map. Wavelet coefficients are then either
set to or shrunk towards zero depending on whether their absolute
values are above or below a chosen threshold. Finally we compute the
inverse wavelet transform of the thresholded coefficients.  We call
this procedure `denoising' (see Donoho 1992 for a discussion of this
term.)  Although the procedure is similar to the one often used with
Fourier coefficients, Donoho (1992) showed that wavelet denoising efficiently
supresses noise while preserving the sharpness of the original
signal. This property makes wavelet denoising appealing for our
applications.

\subsection{Thresholding}
\label{sec.thresh}

`Thresholding' is a prescription for using the sample wavelet
coefficients $\gamma_{j,m}$ to estimate the true wavelet coefficients,
i.e., the wavelet coefficients of the noiseless map.  We use
thresholding functions of the form (Donoho 1992)
\begin{equation}
 \delta(\gamma_{j,m}) =  \left\{
\begin{array}{ll}
\gamma_{j,m} - \tau_j & \mbox{if $\gamma_{j,m} \geq \tau_j$}\\
0               & \mbox{if $|\gamma_{j,m}| < \tau_j$}\\
\gamma_{j,m} + \tau_j & \mbox{if $\gamma_{j,m} \leq -\tau_j$},
\end{array}
\right.
\label{thrule}
\end{equation}
where $\gamma_{j,m}$ is a wavelet coefficient of the map and  
$\tau_j$ is a chosen threshold at scale $j$. A thresholding scheme 
corresponds to a particular prescription for choosing the $\tau_{j}$.
The $\gamma_{j,m}$ are now replaced by the
$\delta(\gamma_{j,m})$ which are used as the ``denoised'' wavelet
coeffecients. 

Given a coarsest resolution $J_o$, Donoho \& Johnstone (1994) suggest
keeping the coefficients $\lambda_{J_o}$ where the signal dominates
the noise and thresholding the rest.  Wavelet thresholding is an
active area of research. Comparing different thresholding methods is
outside the scope of this paper. See Nason (1995) for a survey of
thresholding methods.  We use the {\it SureShrink} procedure of Donoho
\& Johnstone (1995).  Ideally one wants to use the threshold that
minimizes the mean square error (MSE)  
\begin{equation}
{\rm MSE} = \langle\sum_m (\delta(\gamma_{j,m}) - \mu_{j,m})^2\rangle ,
\label{eqrisk}
\end{equation}
where $\mu_{j,m}$ is the mean of the coefficient $\gamma_{j,m}$. 
However, the means $\mu_{j,m}$ are unknown and therefore so is the MSE 
(\ref{eqrisk}). To overcome this difficulty {\it SureShrink}
uses Stein's unbiased estimate of the MSE (see Donoho 1992). 
{\it SureShrink}   
selects a threshold $\tau_j$ at each scale $j$ that is the
best, for rules of the form (\ref{thrule}), in the sense that it
minimizes Stein's estimate of the MSE (\ref{eqrisk}).
The computational cost at each
level is $O(k_j{\rm log}(k_j))$.  
Denoising a map with gaps is done in exactly the same way but using only those 
coefficients based on pixels outside the gaps (see Appendix).

{\it SureShrink} wavelet thresholding is asymptotically minimax
(Donoho \& Johnstone 1995), i.e., it minimizes the worst possible mean
square error of the wavelet coefficients estimates. Under certain
conditions minimax estimators are posterior Bayesian estimates given a
least favorable prior, see Lehmann \& Casella (1998) for a more
complete discussion regarding the relation between minimax and
Bayesian estimators.

\subsection{Noise}
\label{section noise}

Denoising requires information about what constitutes noise and what its
distribution is. When we use wavelets to identify foreground sources,
$F_{i}$, both the cosmological signal $T_{s,i}$ and the noise $n_{i}$ are
assumed to be random components of the noise that contaminates the
wavelet coefficients, i.e., $\sigma^2 = \sigma_{\rm CMB}^2 + \sigma_{\rm
  noise}^2$, and the denoised wavelet coeffecients are identified with
the sources alone. In the second application, when we only suppress
noise in a map, only $n_{i}$ is considered noise, i.e., $\sigma =
\sigma_{\rm noise}$, and the signal component is taken to be the
cosmological CMB component, $T_{s,i}$. These values of $\sigma$ are
used by {\it SureShrink} to compute the estimate of the MSE
(\ref{eqrisk}).

In this paper we assume that pixel noise is uncorrelated and Gaussian.
This was approximately the case for the DMR maps (Lineweaver \etal
1994) but it may not be a reasonable assumption for other
experiments. For cases where the noise is correlated we point out the
following.  If the correlation structure is known then the map can be
transformed to the uncorrelated case; however, depending on the size of
the covariance matrix, such a transformation might be computationally
expensive. Moreover, this transformation is not appropriate when
searching for localized sources because it destroys and widens their
shape. More generally, such a decorrelating transformation no longer
gives a ``sky map'', but a linear combination of sky pixels. Only in
the case of suitably mild correlations can such a decorrelated dataset
still be considered a map per se. One should also note that for the
process of identifying localized sources, when the signal $T_{s,i}$ is
considered part of the noise, the cosmological correlation of
$T_{s,i}$ must also be included in the noise correlation structure.
Depending on the amount of correlation, normalizing only by the
variance (as for uncorrelated noise in Section \ref{secrms}) may be
sufficient. Johnstone \& Silverman (1997) show that {\it SureShrink}
still works well under mild conditions on the correlation structure.
They show that long range correlations in the data are reduced by the
wavelet transform, that there is almost no correlation between
coefficients from different scales, and that Stein's MSE estimate is
still unbiased.

\section{Estimation and Removal of Point Sources with Wavelets}
\label{subloc}
 
We now use the denoising procedure to reduce localized source
contamination in sky maps.  Here $T_{s,i}$ and $n_i$ are assumed to
have zero mean and to be uncorrelated, while $F_i$ is fixed and
unknown.  The goal is to estimate the foreground source field
${\Fb}$. Since the noise and the cosmological model have zero mean
(we assume that the CMB monopole has been subtracted), uncontaminated
denoised wavelet coefficients should be close to zero. Our goal is to
find an estimate $\hat{\Fb}$ of the foreground field ${\Fb}$ 
by thresholding the
wavelet coefficients. We then use a criterion to identify sources in
the field. We suggest masking pixels in identified locations.

To estimate the source field we normalize the coefficients using the
matrix $\Db$ defined by Equation (\ref{Deqn}).  
To estimate the variability of
the wavelet coefficients at each resolution, i.e., the diagonal
elements of $\Db$, we use the median absolute deviation.  This measure
is resistant to outliers and thus less affected by point sources.
Once coefficients have been normalized they are denoised using {\it
SureShrink}.  By assumption, the means in Eq.~\ref{eqrisk} satisfy
$\mu_{j,m}=0$ for all $\gamma_{j,m}$ not contaminated by sources.

\begin{figure} 
\caption{Estimating a source field (top left) from a noisy source
field  
(top right) by thresholding Haar (bottom left) or Daubechies (bottom
right) wavelet coefficients. All sources have identical signal, 
and the s/n is 3, where s/n is defined as (peak amplitude)/(RMS noise). 
Daubechies wavelets yield higher s/n estimates of the source location 
and shape.}
\begin{center}
\begin{tabular}{@{}cc}
\epsfxsize=4cm\epsffile{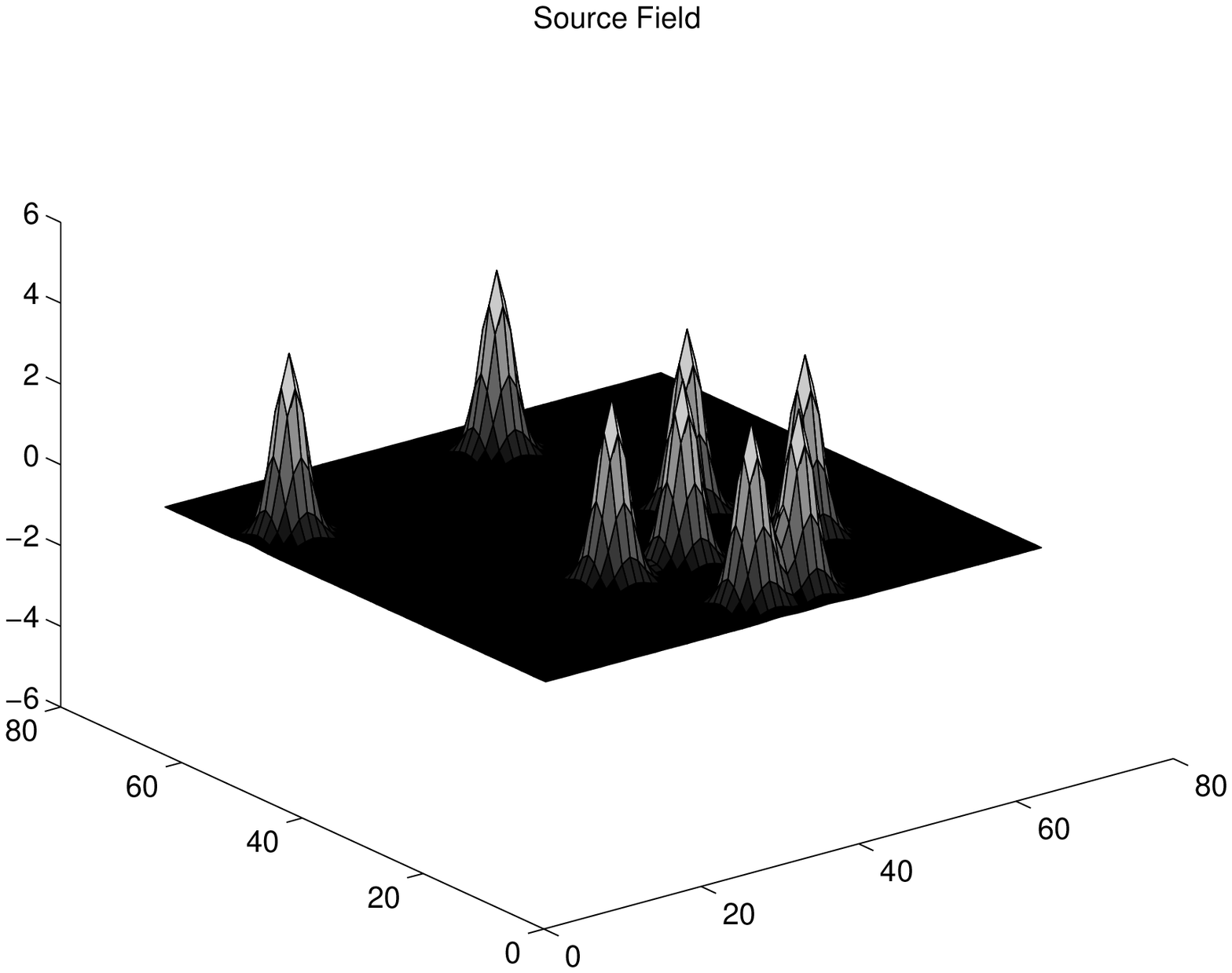} &
\epsfxsize=4cm\epsffile{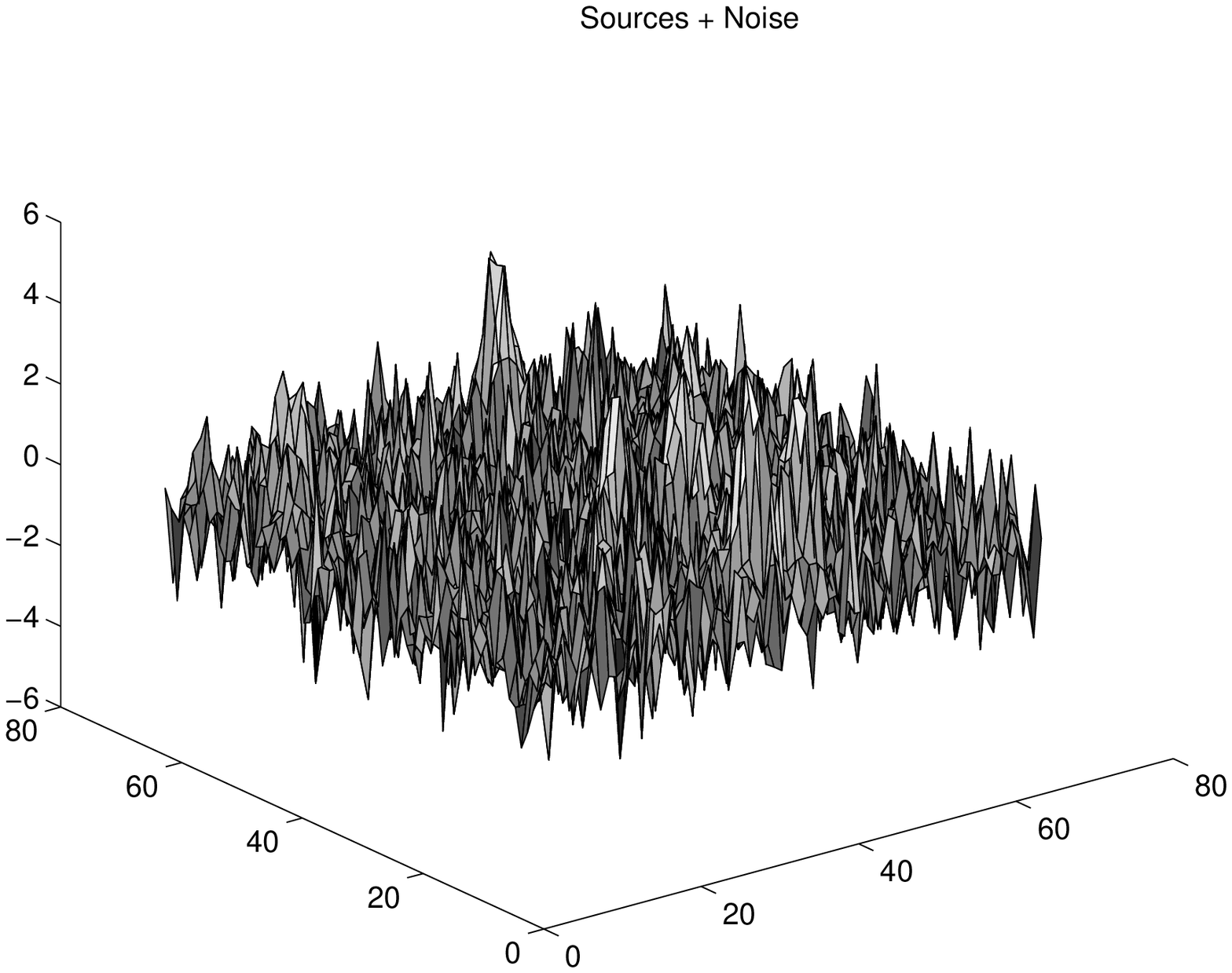} \\
\epsfxsize=4cm\epsffile{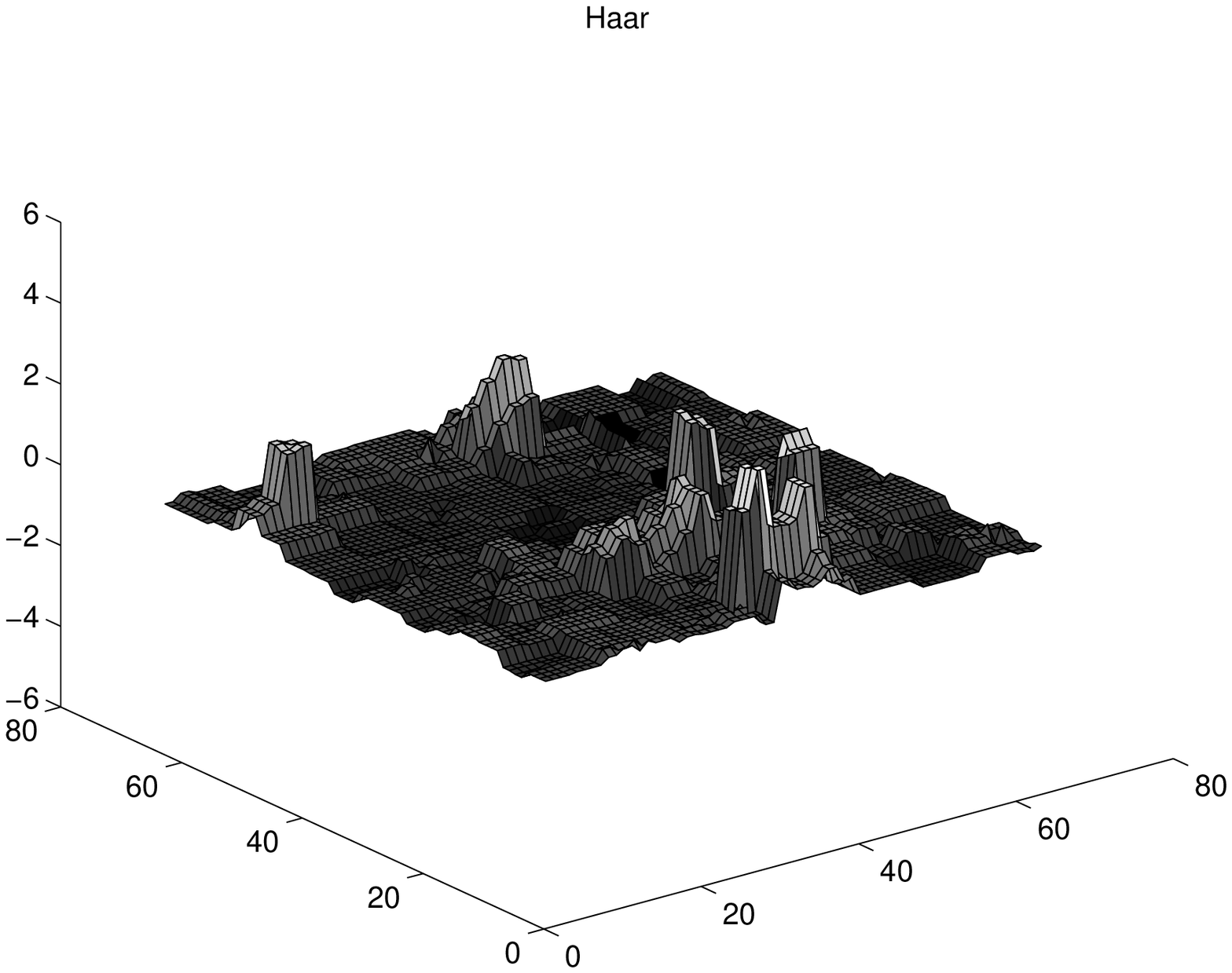} &
\epsfxsize=4cm\epsffile{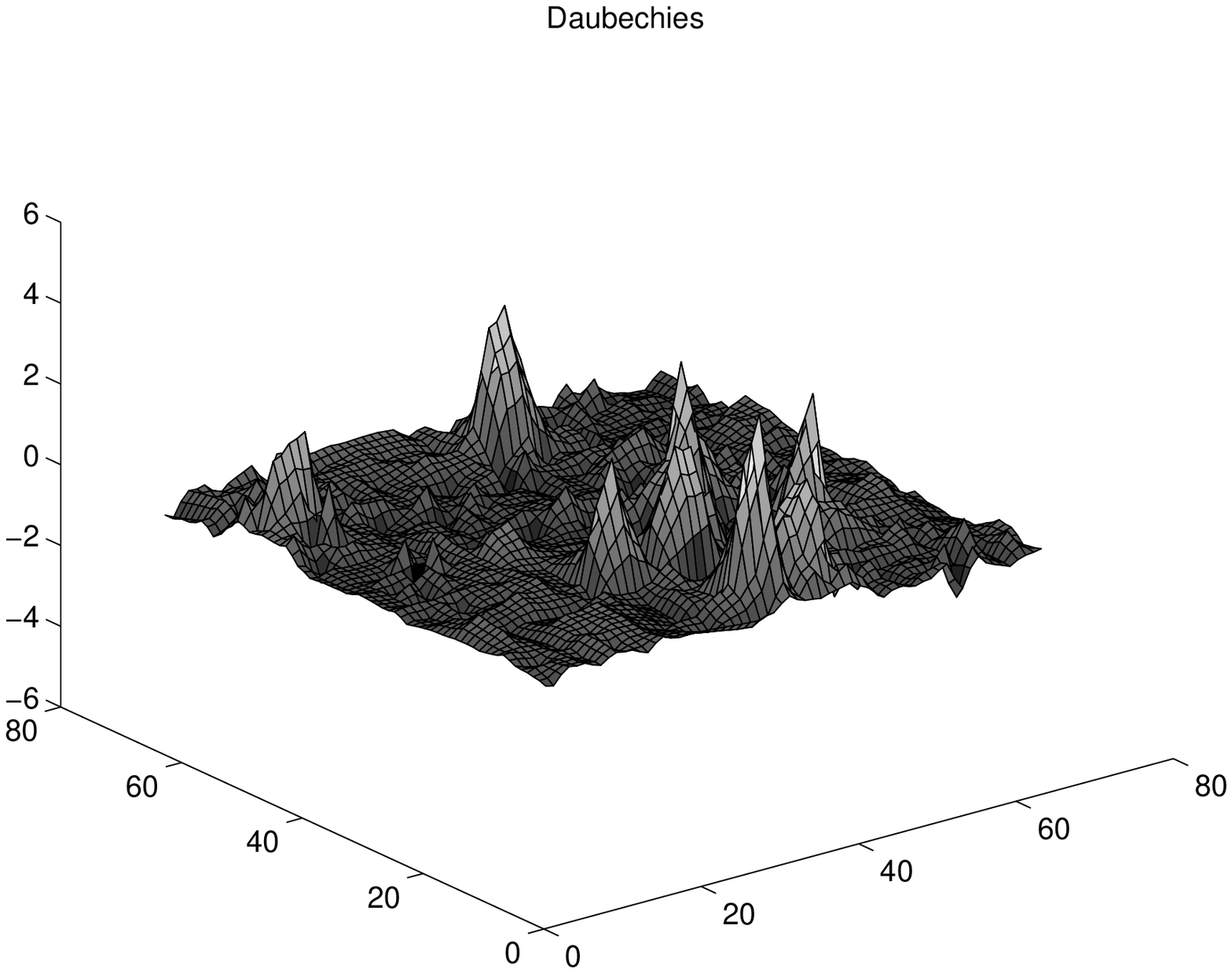} \\ 
\end{tabular}
\end{center}
\label{figloc}
\end{figure}

Using SHW on the whole sky has some drawbacks. First, SHW are not
smooth and therefore do not provide good estimates of smooth peaks of
localized sources. (Localized sources are at least as broad and
smooth as the beam of the experiment.) Also, since local sources
are sparse they are difficult to detect when considering the entire
sky. For noise dominated data {\it SureShrink} uses a large threshold
proportional to $\sqrt{{\rm log}(k_j)}$.  These two problems are
alleviated by taking smaller sky sections where smooth planar wavelets
can be used.  Patches are easily defined through the map pixelization
by taking lower resolution pixels as patches. On each patch we use
tensor products of one-dimensional Daubechies wavelets. (We found that
better results were obtained with four vanishing moments, and that other
wavelet bases do not lead to significantly different results from
those obtained with Daubechies wavelets.)  Note that SHW on small
patches reduce to tensor products of Haar wavelets. Figure
\ref{figloc} shows the original (six sources of the same amplitude)
and original plus white noise 
(s/n = 3, where s/n is the ratio of the peak amplitude to the noise RMS) 
source fields on a patch, and thresholded estimates using planar Haar
and Daubechies wavelets.  Daubechies wavelets recover the peaks of the
sources better than Haar wavelets. Hence we use Daubechies wavelets
for point source identification and removal.  Note also that source
estimates are very close to flat wherever there are no sources. This
minimizes potential systematic effects arising from false source
identification. 

\subsection{Single Source in White Noise}
\label{sswn}

To learn about the denoising process we first apply the procedure to a
single source buried in white noise.  Table \ref{tabhd} (left) shows
the relative error in reconstructing the source. We calculate
RMS$({\Fb}-\hat{\Fb})/{\rm RMS}({\Fb})$ {\it inside} the
source, where `inside' is defined as pixels where ${\Fb}> 1\%$ of
the noise RMS. This measure of relative error is an indication of
power in the source field not accounted for in the estimated
field. The s/n is the ratio of the source peak to the noise RMS. The
size is the ratio between the number of pixels inside the source and
the total number of pixels in the patch in percent. The first and
second rows correspond to Haar and Daubechies wavelets,
respectively. In all cases Daubechies wavelets yield smaller
relative errors.  In Table \ref{tabhd} (right)
we quantify to what extent the wavelet thresholding affects regions
originally not contaminated by the source.  We calculate the ratio
RMS$({\Fb}-\hat{\Fb})/{\rm RMS}({\rm noise})$ {\it outside} the
source.  The table shows that the ripples introduced in the
originally uncontaminated region have an amplitude of less than
20\% of the noise RMS, well within the noise level, and that this
amplitude is very nearly constant with source size.

\subsection{Spectrum of Source Intensity}
\label{sec:sourcespectrum}
We will now attempt to identify and remove a more realistic
distribution of sources from a map. We generate a source field
using a power-law
\begin{equation}
dN(s)\propto s^{-2.5} ds,
\end{equation}
where $N$ is the number of sources and $s$ is signal-to-noise ratio.
(As before, s/n is the ratio of peak amplitude and noise RMS.)  Source
centers are randomly distributed on the patch and only
sources with $s >1$ are included. The power-law exponent, $-2.5$, is appropriate
for a population of sources with a single luminosity distributed
randomly through an infinite, flat, euclidean space.  The total number
of sources in the patch is determined by the choice of different
ratios (total area inside sources)/(total patch area), where `inside'
is defined as in the previous section. The noise is assumed to be
either white or characterized by a Gaussian correlation function with
damping scale as a fraction of the maximum wavenumber.
We used two Gaussian random fields, $G_1$ and $G_2$, with damping scales
0.8 and 0.6, respectively.
The field
$G_1$ has correlations of approximately 12\%, 2\% and 1\% between
first, second and third neighbors, respectively. The correlations for
the second Gaussian field, $G_2$, are 20\%, 3\% and 1\%. 
Rather than using a particular CMB model, we used the Gaussian
correlated noise as a generic model for a correlated
signal. As we will see our results are not sensitive to
the particular correlation structure assumed.

To determine source locations in the thresholded map we perform a
second thresholding on $\hat{\Fb}$.  
This is done by comparing pixel absolute magnitudes with
$k\times ({\rm MAD})$, where MAD is the median absolute deviation of
$\hat{\Fb}$, for different values of the threshold $k$.

\begin{figure} 
\caption{(a) Proportion of correctly detected source pixels that were
above 1$\sigma$ as a function of Area for different thresholds
$k\times$MAD. `Area' is the fraction of pixels covered with sources,
$k$ is an integer and MAD is the mean absolute deviation of the 
denoised map - see text for more details.   (b)
Proportion of pixels outside the source that were incorrectly
identified as source pixels. As a comparison, the dotted lines
correspond to the process of detecting source pixels by setting a
simple $1\,\sigma$ cut on the map.}
\centerline{\epsfxsize=8cm\epsfbox{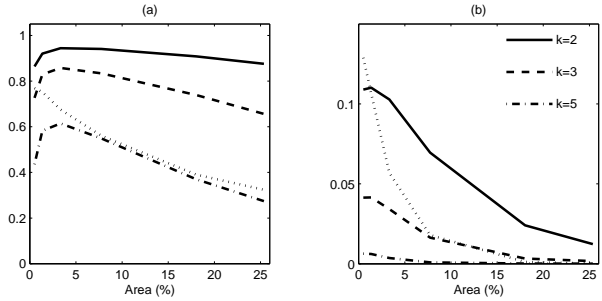}}
\label{detmad}
\end{figure}

\begin{figure}
\caption{(a) Proportion of correctly detected source pixels that were
  above 3$\sigma$ as a function of Area for different thresholds
  $k\times$MAD. `Area' is the fraction of pixels covered with sources,
  $k$ is an integer and MAD is the mean absolute deviation of the
  denoised map - see text for more details.  (b) Proportion of pixels
  outside the source that were incorrectly identified as source pixels.
  As a comparison, the dotted lines correspond to the process of
  detecting source pixels by setting a simple $3\,\sigma$ cut on the
  map.}
\centerline{\epsfxsize=8cm\epsfbox{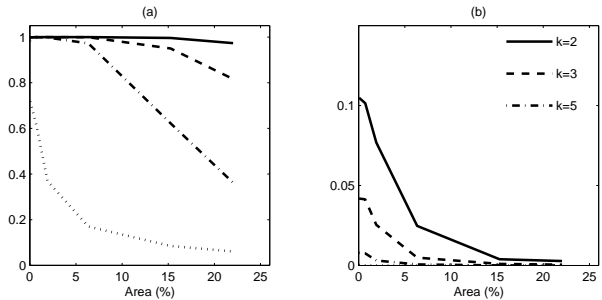}}
\label{detmad2}
\end{figure}

Figures \ref{detmad}, \ref{detmad2} show the results of the
point-source identification process for sources that were above
$1\sigma$ and $3\sigma$, respectively. For each figure, plate (a)
shows the proportion of correctly identified source pixels for
$k=2,3,5$ and plate (b) shows the proportion of pixels outside sources
that were incorrectly identified as source pixels, for the same $k$
values. The area
is defined as the proportion of source pixels above $1\sigma$.  As $k$
increases the probability of correct detection decreases as does that of
incorrect detections. The results in Figures
\ref{detmad}, \ref{detmad2} are for sources in white noise but no
significant difference was observed with Gaussian correlated noise.
Even a Gaussian field with damping scale 0.3 did not show significant
differences.

The figures show that wavelets are extremely effective in identifying
point sources. When we are looking for sources of moderate s/n
(Figure \ref{detmad2}) which cover up to $\simeq 5\% $ of
the patch area, virtually all source pixels are identified regardless of
the value of $k$.  For low s/n sources (Figure \ref{detmad}), a
criterion of $k<3$ identifies more than 80\% of the source pixels.
Comparing Figures \ref{detmad} and \ref{detmad2} we see that for
wavelet thresholding the proportion of false detections does not
change much with the s/n of the source being searched for.  In
addition, simulations show that when there are no sources in
the patch only about 1\% of the pixels in the patch are
incorrectly identified as sources.  To reduce the number of `incorrect
detections' over the entire map, which presumably consists of many
patches, we may choose not to subject every patch to this source
cleaning procedure.  Rather, it may be advantageous to first
calculate the value of $z_{j}$ for the patch, as described in Section
\ref{secrms}, and then proceed with the source cleaning procedure only
on patches with abnormally large values of $z_{j}$ at some scale
$j$. Further analysis and simulations are necessary to optimize the
method.

One can compare the performance of wavelets in identifying point
sources to the performance achieved by a number of alternative
techniques.  A detailed study is in progress. Here we only compare
with the simple practice of removing map pixels whose amplitudes are
larger than $k\sigma$, where $\sigma$ is
the RMS of the noisy source field.  The dotted lines in Figures
\ref{detmad}, \ref{detmad2} correspond to a $1\sigma$ and $3\sigma$
cuts, respectively.  Wavelets achieve a significantly higher
proportion of correct detections, and less errors are made when
searching for low s/n sources.  Interestingly, the $k\sigma$ cut has
fewer errors when searching for high s/n sources. That suggests
that the optimal point source identification procedure may involve a
combination of techniques, with wavelets being the leading tool.
\begin{table*} 
\begin{minipage}{120mm}
\caption{Left: Relative error RMS$({\Fb}-\hat{\Fb})/{\rm
RMS}({\Fb})$ in estimating a field with
a single source embedded in white noise.  
The relative error is calculated using only pixels inside the source,
where 'inside' is defined in the text.  The s/n is the ratio of the
source peak to the noise RMS. The size is the percentage of pixels  
of the patch inside the source. The first and second rows correspond to 
Haar and Daubechies
wavelets respectively. Right: the  ratio of the residual RMS to the
noise RMS outside the source.}
\label{tabhd}
\begin{tabular}{@{}lcr}
{\begin{tabular}{@{}cccccc}
  &    \multicolumn{5}{c} {size (\%)}\\
s/n  &  18 & 11 & 5 & 3 & 1 \\
\hline
2 & .33 & .64 & .83 & 1.02 & 1.34 \\
  & .26 & .53 & .73 & .93 & 1.28 \\
4 & .23 & .41 & .52 & .63 & .81 \\
  & .17 & .29 & .40 & .51 & .74 \\
6 & .19 & .32 & .42 & .52 & .67 \\
  & .13 & .22 & .31 & .41 & .57 \\
\end{tabular}} & $\,\,\,$ &
{\begin{tabular}{@{}ccccc}
\multicolumn{5}{c} {size (\%)}\\
  18 & 11 & 5 & 3 & 1 \\
\hline
   .11  & .11   & .11    &  .11  &  .10    \\
  .12   & .11   & .11    &  .11  &  .10 \\
  .15   & .15   & .15    &  .15  &  .15    \\
  .16   & .16   & .16    &  .16  &  .15 \\
  .17   & .16   & .16    &  .16  &  .16    \\
  .17   & .16   & .16    &  .16  &  .16  \\
\end{tabular}} \\
\end{tabular}
\end{minipage} 
\end{table*}

\section{Map Denoising and Compression}
\label{secmap}
In this section we apply wavelet denoising with the goal of
compressing information and reducing noise for visualisation of
CMB maps. Denoised maps are useful for displaying or transmitting
compressed map files.

Here we assume that the CMB sky, ${\Tb}_s$, is fixed and that the
random field we are trying to isolate and remove is the instrumental
noise.  Hence, only the noise variability is included in the matrix $\Db$
(Eq. \ref{Deqn}), and it is estimated using only the wavelet
coefficients at the highest resolution where we expect little
structure. Alternatively, $\Db$ could be estimated using the
difference of two independent maps (as with the $A-B$ DMR maps).
This time {\it
SureShrink} determines the threshold $\tau_j$ that minimizes
(\ref{eqrisk}) with $\mu_{j,m}$ being the unknown wavelet coefficients
of the fixed sky ${\Tb}_s$.  The denoised estimate of
${\Tb}_s$ is then $\Wb^{-1}\delta(\gamma)$, the inverse transform
of the thresholded coefficients. The computational cost is $O(N{\rm
log}(N))$.  Note also that the estimate of the underlying temperature
field achieved in this process is model independent. In other words,
it did not require a cosmological model.

In other work Wiener filtering has been used for
denoising CMB maps (Bunn \etal 1996).
The Wiener filter estimate of
${\Tb}_s$ is $\Cb {\db}$, where the matrix $\Cb$ is determined by
assuming a cosmological model for ${\Tb}_s$ and minimizing
$\langle||\Cb {\db}- {\Tb}_s||^2\rangle$. The process takes
$O(N^3)$ operations and depends on the fiducial model chosen. 
Thus, while Wiener filtering provides a
posterior mean estimate given a prior cosmological model, wavelet
thresholding provides an estimate of the actual realization of the
unknown field without assuming a cosmological model. 
The Wiener filter can be defined as the maximum-probability signal
contribution given the data, the underlying signal power spectrum and
noise correlation, and assuming Gaussianity for both signal and noise.
{\it SureShrink} wavelet thresholding, on the other hand, is
aymptotically minimax (see Section \ref{sec.thresh} and  
Donoho \& Johnstone 1995).
 
\begin{figure} 
\caption{Top: a simulated, signal only,
CMB map based on a standard CDM model. There are about 98000 pixels
in the map.
Middle: the signal map with noise added. The s/n in this map is
$\approx .7$.
Bottom: The noisy map after denoising with {\it SureShrink}.
Only 14\% of the wavelet coefficients were used, and
the denoising process took less than 6 seconds on an Alpha
500/266 workstation.}
\begin{center}
\begin{tabular}{cc}
\multicolumn{2}{c}{\epsfxsize=7cm\epsffile{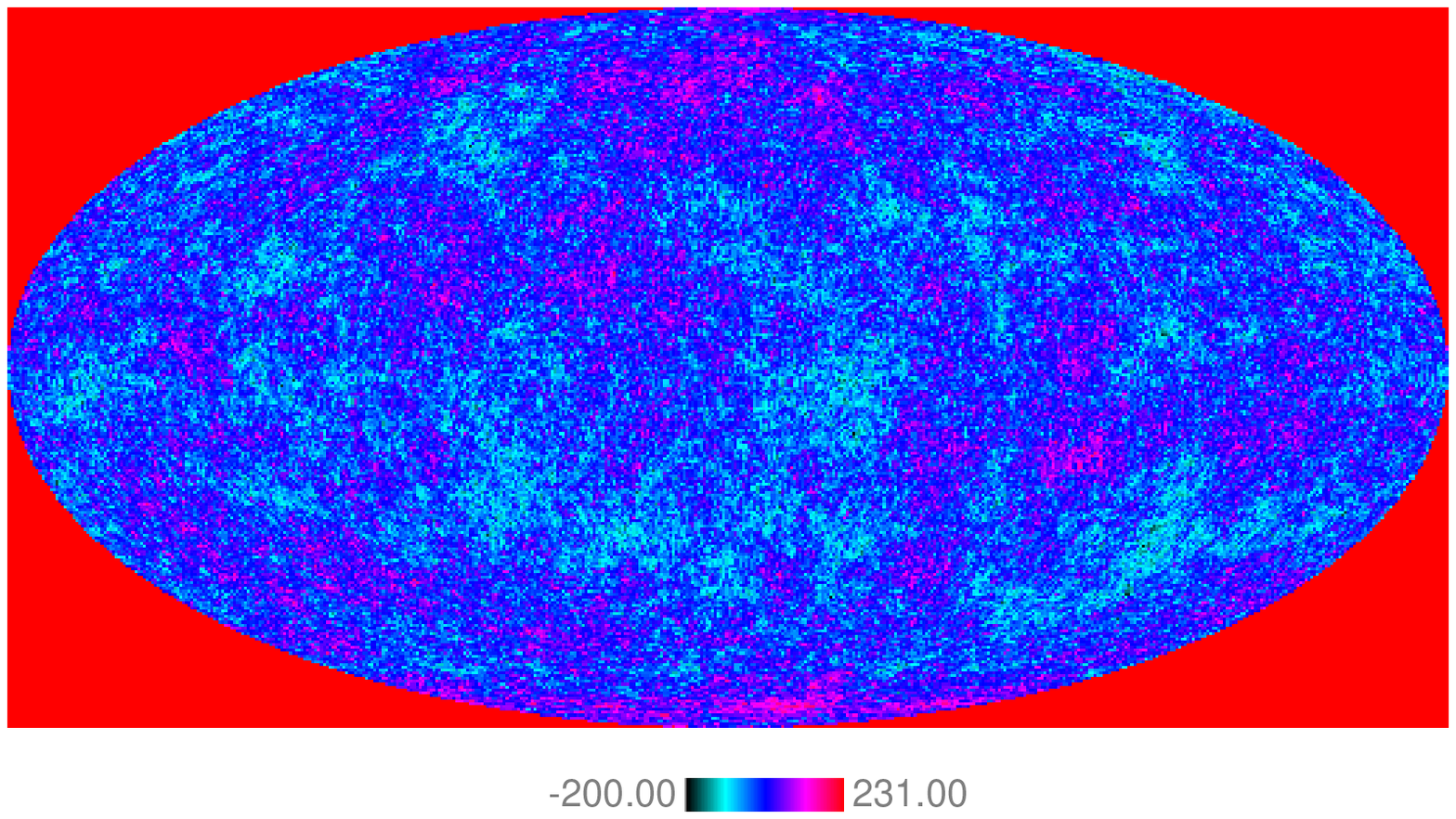}}\\
\multicolumn{2}{c}{\epsfxsize=7cm\epsffile{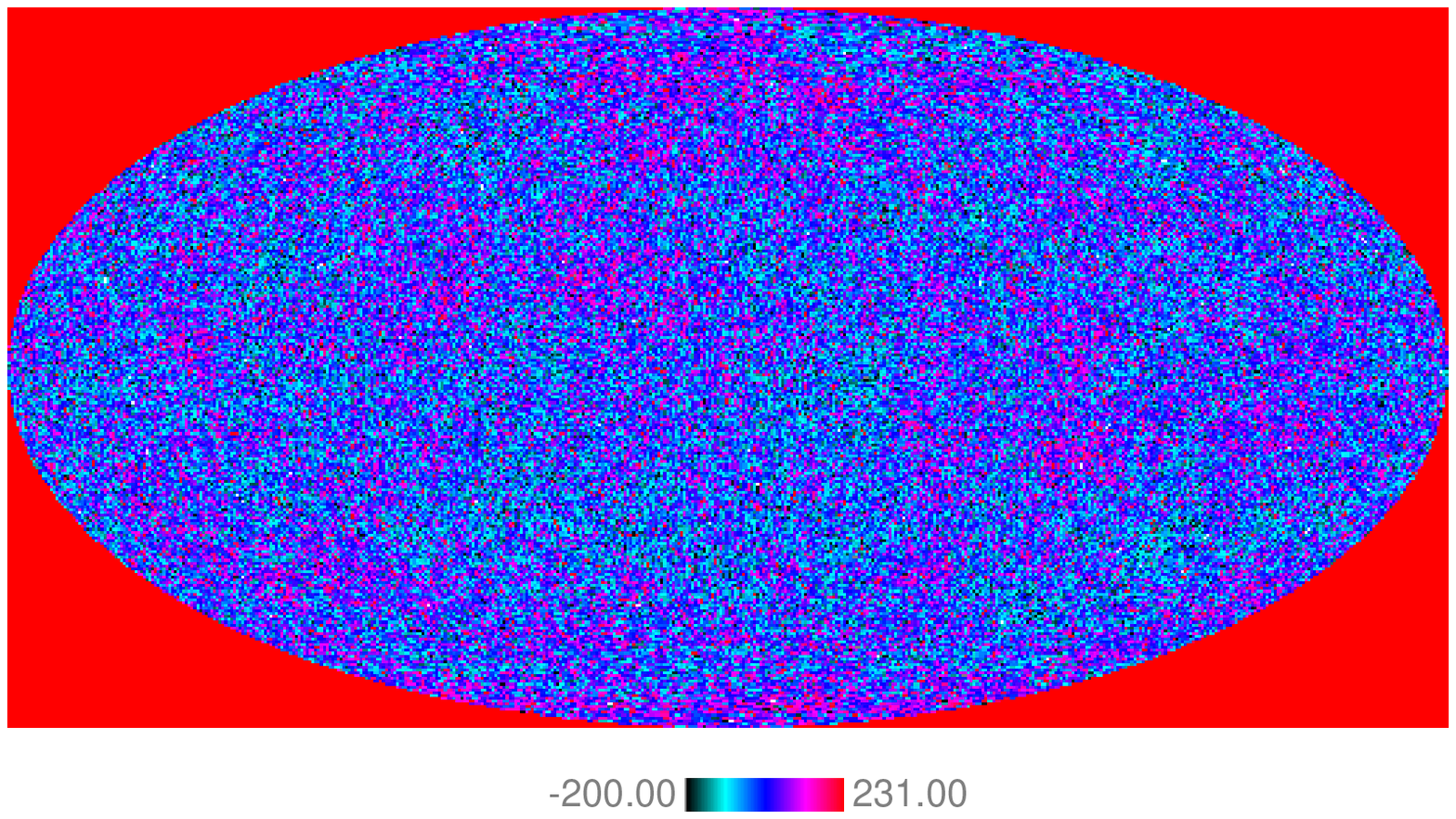}}\\
\multicolumn{2}{c}{\epsfxsize=7cm\epsffile{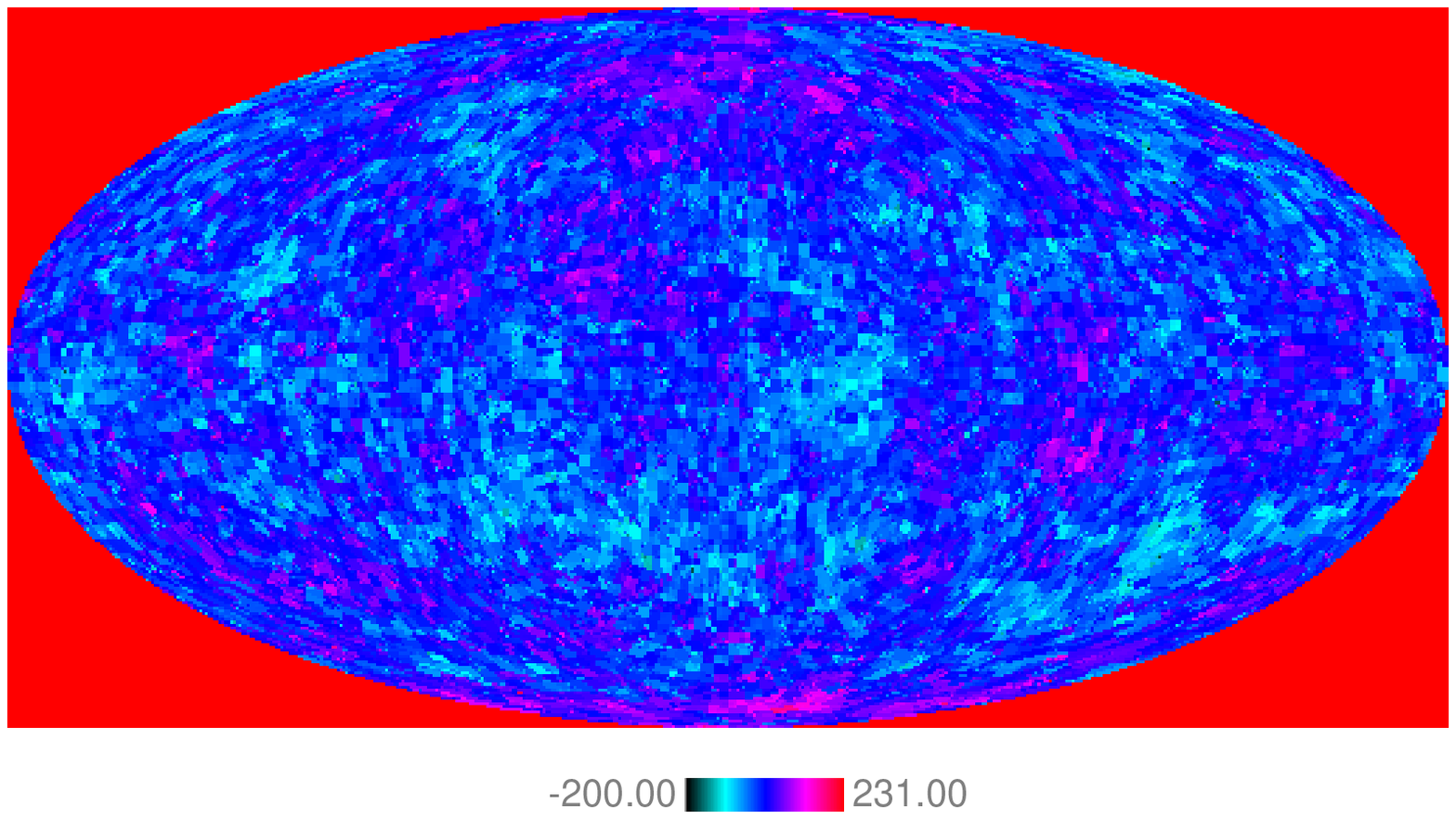}}\\
\end{tabular}
\end{center}
\label{maps}
\end{figure}
Figure \ref{maps} shows the denoising of a simulated standard CDM map
with $\Omega_b=.05$, $\Omega=1$, $H_0=50$~km/s/Mpc. The
signal to noise in the signal plus noise map (middle) is $\approx .7$, where
the signal to noise is the ratio of the noise RMS to the signal RMS. 
There are 98304 pixels in the map. 
The denoised map (bottom) achieved a 35\% noise reduction, as measured
by the ratio RMS(true map - denoised map)/(noise RMS), 
using only $14\%$ of the 98304 wavelet coefficients. 
It also has only 14\% of the original number of pixels, 
but pixels have a variety of sizes, adapted to 
the local structure as identified by the thresholding procedure.
The entire denoising process took less than 6 seconds on an Alpha 500/266
workstation.  The compression rate that denoising can achieve depends
on the signal to noise. As it increases more coefficients have to be
used.  Table 3 shows the noise reduction and compression rate for
different signal to noise ratios using the same standard CDM model.

Wavelet denoising is a flexible procedure in that it
allows one to
choose the threshold level which will achieve a 
particular compression rate regardless
of the denoising achieved.  In our case the thresholds were chosen to
minimize an estimate of the mean square error (\ref{eqrisk}) but this
may not be the optimal procedure.  Depending on the goal at hand one
may be able to devise a thresholding procedure tailored for a
particular task.
 
\begin{table} 
\caption{ Noise reduction and compression achieved
by {\it SureShrink} denoising of a standard CDM model for different
signal to noise ratios.}
\label{tbl3}
\begin{tabular}{@{}ccc} 
 signal/noise  &  RMS(residuals)/RMS(noise)   & \% coefficients used  \\ \hline
 .9     &       .82     &       33 \\
.7      &       .65     &       14 \\
.6      &       .57     &       7 \\
\end{tabular} 
\end{table}

\section{SHW and Cosmological Information}
\label{secparam}

It has become traditional to consider the cosmological information from
CMB datasets in the context of likelihood functions, usually expressed
in pixels or in the spherical harmonic domain. Since no information is
lost by transforming to wavelet space we can ask whether it is better to
work in wavelet domain when the goal is to estimate $C_{\ell}$s and
cosmological parameters. 

Assuming Gaussianity of both the cosmological
signal and the noise, the log-likelihood of the data given a model with
covariance matrix $\Mb$ is
\begin{eqnarray}
 -2{\rm ln}\, {\cal L}(p_i | \db) & = &{\rm ln}\,|\Mb| + \db^t \Mb^{-1} \db
\nonumber \\ 
& = & {\rm ln}\,|\Sb_\gamma| + \gamma^t \Wb^{-t}  \Mb^{-1}\Wb^{-1} \gamma
\label{likeq}
\end{eqnarray}
(up to an additive constant) where $p_i$ are the cosmological
parameters and $\gamma = \Wb\db$ is the wavelet transform of the data $\db$.
$\Mb=\Cb+\Sb_{\rm noise}$ is the sum of contributions from the
signal and the noise covariance. In the second equality we have
transformed from the map pixel basis to the wavelet domain (see also
Eq.~\ref{covgamma}).  To estimate the $p_i$ it is enough to consider
the eigenmodes of $\Cb$ which contribute the most to ${\cal L}$ (this
is the Karhunen-Loeve or ``signal-to-noise eigenmode''
transformation, Bond 1995).  This simplifies the likelihood by compressing the
data. The reduced likelihood can be used to estimate cosmological
parameters. 

Can we apply the same technique on the thresholded wavelet
coefficients?  Because the thresholding procedure is nonlinear the
distribution of the {\it denoised} $\gamma = \delta(\gamma)$ is no
longer Gaussian and therefore Eq.~\ref{likeq} written in terms of
$\delta$ is no longer the correct log-likelihood.  It is
computationally expensive to compute the correct likelihood function
of the thresholded coefficients for each plausible cosmological
model. We can still maximize Eq.~\ref{likeq}, with respect to $p_i$,
to estimate the parameters but the perfomance of this procedure is
still to be investigated.

The difficulties due to the non-linearity of the wavelet thresholding 
process can perhaps be
resolved by treating the process as a linear operation. That is, we
consider the thresholding operation of Eq.~\ref{thrule} as if the
thresholds $\tau_j$ are fixed beforehand. Then, the new wavelets
coeffecients $\delta(\gamma_{j,m})$ can be considered as linearly
related to the old coeffecients: $\delta(\gamma_{j,m}) =
\alpha_{jm}\gamma_{j,m}$. This is similar to, for example, Wiener
filtering, where each Fourier mode is modified by some factor
$0\le\alpha_k\le1$ (although in that case the process is strictly
linear since the $\alpha_k$ do not depend upon the data). Thus, the
denoised wavelet coeffecients can be considered a much smaller linear
rendition of the original data. The correlation structure may indeed
be much more complicated than the raw data, but because there may be
many fewer coefficients, the $O(N^3)$ operations necessary for
manipulating the likelihood are much faster. This procedure is under
further investigation.

Since working with the denoised coefficients seems problematic one
could use the likelihood without denoising the coefficients. But then
nothing seems to be gained by working in wavelet domain since the
quadratic term in (\ref{likeq}) does not split into independent
components of different scales---the second equality above is not
easier to compute than its version in the pixel domain.  Consider, for
example, a quadratic estimator $\lambda_J^t \Eb \lambda_J$ of
$C_{\ell}$ (e.g., Tegmark 1997; Bond, Jaffe \& Knox 1998).  Since the
wavelet transform acts as a band-pass filter, one might hope to reduce
leakage and decrease computational cost by considering only high
resolution coefficients to estimate small-angular-scale (large $\ell$)
$C_\ell$, splitting $\lambda_J^t \Eb \lambda_J$ as in (\ref{rmsdec})
and taking only high resolution coefficients. There are three
problems: (1) Spherical harmonics have components in the lower wavelet
scales (Table 1) that need to be estimated; (2) Wavelet coefficients
$\lambda_j$ are orthonormal with respect to the identity but not with
respect to any matrix $\Eb$. In general the cross terms in a quadratic
estimator using a decomposition of the form (\ref{rmsdec}) do not
cancel; (3) The number of coefficients at each scale grows
exponentially; using only the highest resolutions yields little
compression. For example, 93\% of the coefficients of a resolution 10
map belong to the two highest resolutions. Because of these problems,
the estimator (like the likelihood function itself above) does not
appear to be any easier to compute in the wavelet domain.
 
\begin{figure} 
\caption{Histograms of the $z_4$ statistic for the same kind of simulations
  as in Figure \ref{dmrzs} but with spectral index $n=0.5$ (instead of
  $n=1$), where $n$ is the spectral index of the primordial desnity
  fluctuation power spectrum which in turn gives $\ell(\ell+1)C_\ell
  \propto \ell^{n-1}$. The distribution of $z_4$ under this model is
  clearly different from that of the $n=1$ model (Figure \ref{dmrzs}).
  The dashed line indicates the value of $z_4$ for the DMR 53+90 GHz
  map. }
\centerline{\epsfxsize=7cm\epsfbox{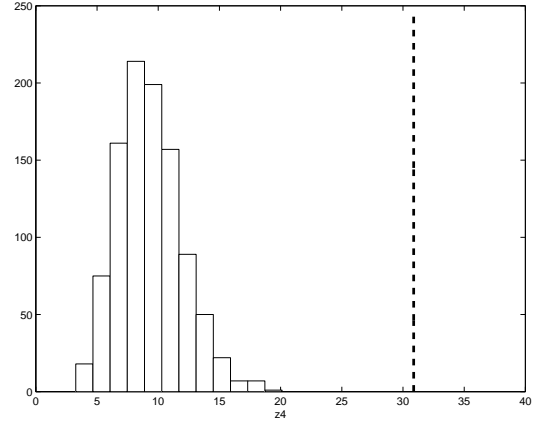}}
\label{dmrzsnp5}
\end{figure}

Another approach is to consider using the wavelet coefficients
themselves as intermediate parameters in the process of cosmological
parameter estimation.  A homogeneous random field is characterized by
its power spectrum $C_{\ell}$, which is in turn parametrized by the
cosmological parameters. In principle cosmological parameters can be
estimated by fitting to the $C_{\ell}$.  Power in the wavelet domain
provides some information about cosmological parameters. For example,
Figure \ref{dmrzsnp5} shows the distribution of the normalized power at
resolution 4 ($z_4$) for the same large-angular-scale models of
Figure \ref{dmrzs} but now using $n=0.5$ (instead of $n = 1$), where
$n$ is the spectral index of the primordial density fluctuation power
spectrum which in turn gives $\ell(\ell+1)C_\ell \propto \ell^{n-1}$.
Comparing this histogram to the one in Figure \ref{dmrzs} it is obvious
that the distribution of $z_4$ can discriminate between the two models.
However, even future high resolution CMB maps will only be of
resolutions $j\leq 10$, and only the highest resolutions $j\geq 7$ will
be most useful for extracting cosmological parameters . Using the
wavelet power at scales $j=7,8,9,10$ will not yield enough degrees of
freedom to fit 11 cosmological parameters. This paucity of information
in the wavelet domain is because the number of independent wavelet power
coefficients (the $z_j$s) scales only as the logarithm of the number of
pixels. In contrast the number of independent power spectrum
coeffecients $C_\ell$ (which completely determine the properties of the
homogeneous and isotropic Gaussian CMB field) scales as the
square root of the number of pixels.  Using the wavelet power as we have
defined it here will not be sufficient.

We could instead consider abandoning both $C_\ell$ and the wavelet power
defined here in exchange for a spectrum more localized in both location
and scale. The question is whether we can find a spectral representation
of homogeneous random fields on the sphere in terms of localized
functions. However, it is easy to show 
that there is no smooth basis
$\{\Psi_{i,j}\}$, with local support decreasing with $j$, with the
property that any homogeneous random field on the sphere can be written
as $\sum_{i,j} \alpha_{i,j} \Psi_{i,j}$ where $\langle \alpha_{i,j},
\alpha_{i',j'} \rangle = \sigma_j^2 \delta_i^{i'} \delta_j^{j'}$. That
is, any other ``power spectrum'' we define in terms of locally supported
functions will have an extended ``window function'' in $\ell$, as with
the wavelet coeffecients of Section~\ref{secrms}; the usual $C_\ell$
spectrum is still the most natural choice.

\section{Summary}
\label{secsum}

Wavelets provide a useful tool to investigate local structure in maps.
We used SHW to define a local measure of power that is equivalent to a
normalized local angular scale decomposition of the map RMS. This
measure can be used to compare power in different regions in the
sky. Similar measures can be defined with other bases of orthogonal
wavelets. SHW have the advantage of being easily implemented given any
map pixelization scheme.  SHW can also be used to denoise and compress
maps without doing any diagonalization or inversion of matrices.
We denoised and compressed a 100,000 pixel CMB map
by a factor of $\sim 10$ in 5 seconds achieving a noise
reduction of about 35\%. In contrast to Wiener filtering the
compression process is model independent. The total computational
cost of wavelet transforming and thresholding a region with $N$ pixels
is $O(N{\rm log}(N))$. 

Since wavelets are locally supported they can be used to reduce
localized source contamination in maps.  We applied planar Daubechies
wavelets for the identification and removal of points sources from
small planar-like sections of sky maps. Our technique successfully
identified most point sources which are above $3\sigma$ and more than
80\% of those above $1\sigma$.  In addition, wavelet thresholded
estimates of source fields introduce little structure in
uncontaminated regions.

We have concentrated on local source detection and map compression.
We did not find useful direct applications of spherical wavelets to
the estimation of angular power spectra or cosmological parameters.
Ideally we would like to have a wavelet basis which approximately
diagonalizes the covariance matrix of the comological model as well as
the noise covariance matrix, thus simplifying the likelihood
function. This is not the case for SHW. But in the future there may be
hope for spherical wavelets that are more compatible with spherical
harmonics.  See for example Freeden \& Windheuser (1997), and
references therein. They define smooth spherical wavelets that combine
a spherical harmonic expansion for the low order terms and wavelet
expansions for the small scale structure.  However, they do not yet
have a fast wavelet transform.

In Section \ref{sec.thresh} we pointed out that {\it SureShrink}
wavelet thresholding provides posterior estimates for a priori
distributions of the wavelet coefficients that give the largest mean
square error. Although we have not used them there are also Bayesian
thresholding methods that can be used to include more informative
prior information on the wavelet coefficients.  See Abramovich \etal
(1997) and references therein.  The development of wavelet tools in
statistics is a very active field of research. For a review see
Silverman (1999) as well as the other articles in the same volume.

The computer code used for work presented in this paper
will soon be freely available at \\
http://cfpa.berkeley.edu/combat.\\
\\
{\bf Acknowledgements}.
This work was done under the auspices of the COMBAT collaboration
supported by the NASA AISR program, grant NAG5-3941. We are grateful to
L.~Cay\'{o}n, D.~Donoho , P.~Ferreira and P.~Stark for helpful
suggestions.  Simulations of CMB maps were simplified thanks to
Seljack \& Zaldarriaga's CMBfast and E.Scannapieco's CMAP
(http://cfpa.berkeley.edu/combat).  We thank G. Smoot for
allowing access to his Alpha cluster. S.H. was
partially supported by NASA grant NAG5-3941, A.J. by NASA LTSA NAG5-6552,
and L.T. by grants NAG5-3941 and DMS-9404276.
Part of this work uses the {\it Wavelab} package, available from
http://www-stat.stanford.edu/wavelab/.

\appendix
\section{Haar Wavelets for the COBE Pixelization}
\label{sechaar}

Spherical Haar wavelets (Sweldens 1995, Schroeder \& Sweldens 1995)
are a generalization
of planar Haar wavelets.  
To define SHW we need a hierarchical
pixelization scheme. One example of such is the COBE sky cube (CSC)
pixelization where the   
surface of the unit
sphere is divided into $6\cdot 4^{(j-1)}$ approximately equal
area pixels at resolution $j$. 
Each pixel $k$ at resolution $j$, $S_{j,k}$, has four 
children, $S_{j+1,k_1},...,S_{j+1,k_4}$, at resolution $j+1$. 
Let $\Ksc (j)$ be the pixel numbers corresponding to resolution
$j$.  

The functions that model the main features of the data at each scale 
are scaling functions, and the functions
that represent the details of the data not captured by the scaling functions
are the wavelets.
The scaling functions of the SHW are defined as
\( \varphi_{j,k}(\eta) = 1 ,\) if $\eta \in S_{j,k}$ and 0 otherwise.
Define $V_j \subset L_2$ as the closed subspace generated by
the $\varphi_{j,k}$:
\(V_j = {\rm clos}\{\varphi_{j,k} | \,\,k\in\Ksc (j) \} .\)
For example, a DMR sky map is an element of $V_6$. 
By definition
$V_{j} \subset V_{j+1}$. The SHW are defined as an 
orthogonal basis for the orthogonal complement, $W_j$, of $V_j$ in
$V_{j+1}$. 
By construction, for any chosen $J$,
\( V_J = V_j \bigoplus_{i=j}^{J-1} W_i. \)
For example, a resolution 10 map is a sum of a coarser
resolution 9 map plus a function from $W_9$ encoding the details not captured
by the coarser map, or 
a resolution 8 map plus detail functions from $W_8$ and $W_9$.
The functions that capture the leftover details are combinations of the 
wavelets at different resolutions: there are three wavelets, $\psi_{j,m_1}
\psi_{j,m_2},\psi_{j,m_3}$, associated to each pixel
$S_{j,k}$. Let $\mu_{j,k}$ be the area of pixel $S_{j,k}$ and 
$\Lsc (j,k) = \{m_o(k),m_1(k),m_2(k),m_3(k)\}$ 
be the pixel numbers of its four children, then the three
wavelets for pixel $S_{j,k}$ are
\[
\psi_{j,m_1}  =  \frac{\varphi_{j+1,m_o}+\varphi_{j+1,m_3}}
{2\mu_{j+1,m_o}+2\mu_{j+1,m_3}} - \frac{\varphi_{j+1,m_1}+\varphi_{j+1,m_2}}
{2\mu_{j+1,m_1}+2\mu_{j+1,m_2}}\]
\[
\psi_{j,m_2} =  \frac{\varphi_{j+1,m_1}}{2\mu_{j+1,m_1}} -
\frac{\varphi_{j+1,m_2}}{2\mu_{j+1,m_2}}\]     
\[ \psi_{j,m_3}   =  \frac{\varphi_{j+1,m_o}}{2\mu_{j+1,m_o}} -
\frac{\varphi_{j+1,m_3}}{2\mu_{j+1,m_3}} \]
 
The wavelets $\{\psi_{j,m_i(k)} |\,\,
i=1,2,3;\,\,k\in \Ksc(j) \}$ form a basis for $W_j$. In the limit, for
any choice of $J_o$,  
$\{\psi_{j,m}, j\geq j_0\}\cup\{\phi_{j_0,k} |\,k\in \Ksc (j_0)\}$
is an orthogonal basis for the space of integrable functions on the sphere.
For example, take a function $\Tb$ (e.g., a sky map) at a finest resolution
level $J$ (i.e., $ {\Tb} \in V_J$) and choose a starting resolution
$J_o$, then $ {\Tb}$ can be written as
\begin{equation}
  T(i) = 
\sum_{k\in \Ksc(j_0)} {\lambda}_{j_0,k}\varphi_{j_0,k}(i) +  
\sum_{j=j_0}^{J-1}\sum_{m} \gamma_{j,m}\psi_{j,m}(i).
\label{expansion}
\end{equation}
The first sum corresponds to the lowest resolution map 
and the second one
to the details from the higher resolutions.
The coefficients in the expansion (\ref{expansion}) define the wavelet
transform $\Wb {\Tb}$ of $ {\Tb}$
\begin{eqnarray}
\Wb {\Tb} & = & \vec{\gamma} \nonumber  
 (\vec{\lambda}_{j_o}, \vec{\gamma}_{jo},\cdots,
\vec{\gamma}_{J-1})^t \nonumber \\
 & = & (\{\vec{\lambda}_{J_o,k}, k\in \Ksc(J_o)\},
\{\vec{\gamma}_{j,m}, \nonumber \\
 &  & m\in\Msc(j),J_o\leq j\leq J-1\})^t . 
\label{coefs}
\end{eqnarray}
No linear system needs to be solved in order to compute the 
components of $\vec{\gamma}$, they are obtained recursively starting from the 
finest resolution coefficients $\vec{\lambda}_{J}= {\Tb}$ 
\[ \lambda_{j,k}=\sum_{l\in\Ksc(j+1)} \tilde{h}_{j,k,l} 
\lambda_{j+1,l},\,\,\,\mbox{and }  
\gamma_{j,m}=\sum_{l\in\Ksc(j+1)} \tilde{g}_{j,m,l}
\lambda_{j+1,l} . \]
To reconstruct a function given the wavelet coefficients we have 
a recursive inverse transform  
\[ \lambda_{j+1,l}=\sum_{k\in\Ksc(j)} h_{j,k,l}\lambda_{j,k} + 
\sum_{m\in\Msc(j)} g_{j,m,l}\gamma_{j,m}. \]
The coefficients for the forward and inverse transforms defined above are
  
\[ \tilde{h}_{j,k,l} = \left\{ 
\begin{array}{ll}
\frac{\mu_{j+1,l}}{\mu_{j,k}} & \mbox{if $l\in \Lsc (j,k)$}\\
0                             & \mbox{otherwise}
\end{array}
\right. 
\]
\[
h_{j,k,l} = \left\{
\begin{array}{ll}
1 & \mbox{if $l\in \Lsc (j,k)$}\\
0 & \mbox{otherwise}
\end{array}
\right.
\]
\[ \tilde{g}_{j,m,l} = \left\{
\begin{array}{ll}
\int_{S_{j+1,l}} \tilde{\psi}_{j,m}\,\,d\mu & \mbox{if $l\in 
\Lsc (j,k)$}\\
0                             & \mbox{otherwise}
\end{array}
\right.
\]
\[
g_{j,m,l} = \left\{
\begin{array}{ll}
\frac{1}{\mu_{j+1,k}}\int_{S_{j+1,l}} \psi_{j,m}\,\,d\mu & \mbox{if 
 $l\in \Lsc (j,k)$}\\
0  &  \mbox{otherwise,}
\end{array}
\right.
 \]
where $\tilde{\psi}_{j,m}$ is the dual of ${\psi}_{j,m}$.

In order to avoid biasing analyses of CMB skies, data contaminated by
Galactic emissions or other identified foreground sources are usually
discarded. To compute a wavelet transform of data with gaps, we
discard those pixels at the lowest resolution $J_o$ which are inside
the gaps.  Only the wavelet coefficients correponding to descendants
of pixels outside the gaps are used. Note that the scaling and wavelet
functions corresponding to these selected pixels form an orthogonal
basis for the sky with gaps. This is due to the local support of the
functions and to the vanishing of the integral of $\psi_{j,m}$ over
any pixel $S_{j,k}$.

\end{document}